\documentclass[fleqn,usenatbib]{mnras}

\usepackage[T1]{fontenc}

\DeclareRobustCommand{\VAN}[3]{#2}
\let\VANthebibliography\thebibliography
\def\thebibliography{\DeclareRobustCommand{\VAN}[3]{##3}\VANthebibliography}

\usepackage{graphicx}	
\usepackage{adjustbox}  
\usepackage{amsmath}	
\usepackage{amssymb}	
\usepackage{wasysym}    
\usepackage{bm}         
\usepackage{xspace}     
\usepackage{etoolbox}
\usepackage{float}
\makeatletter 
  \patchcmd{\NAT@citex}
    {\@citea\NAT@hyper@{%
      \NAT@nmfmt{\NAT@nm}%
      \hyper@natlinkbreak{\NAT@aysep\NAT@spacechar}{\@citeb\@extra@b@citeb}%
      \NAT@date}}
    {\@citea\NAT@nmfmt{\NAT@nm}%
    \NAT@aysep\NAT@spacechar\NAT@hyper@{\NAT@date}}{}{}

  \patchcmd{\NAT@citex}
    {\@citea\NAT@hyper@{%
      \NAT@nmfmt{\NAT@nm}%
      \hyper@natlinkbreak{\NAT@spacechar\NAT@@open\if*#1*\else#1\NAT@spacechar\fi}%
        {\@citeb\@extra@b@citeb}%
      \NAT@date}}
    {\@citea\NAT@nmfmt{\NAT@nm}%
    \NAT@spacechar\NAT@@open\if*#1*\else#1\NAT@spacechar\fi\NAT@hyper@{\NAT@date}}
    {}{}
\makeatother
\usepackage{newtxtext,newtxmath} 

\newcommand\orcid[1]{\href{http://orcid.org/#1}{\adjustbox{trim={-.15\width} {0\height} {-.15\width} {0\height},clip}{\includegraphics[height=10pt]{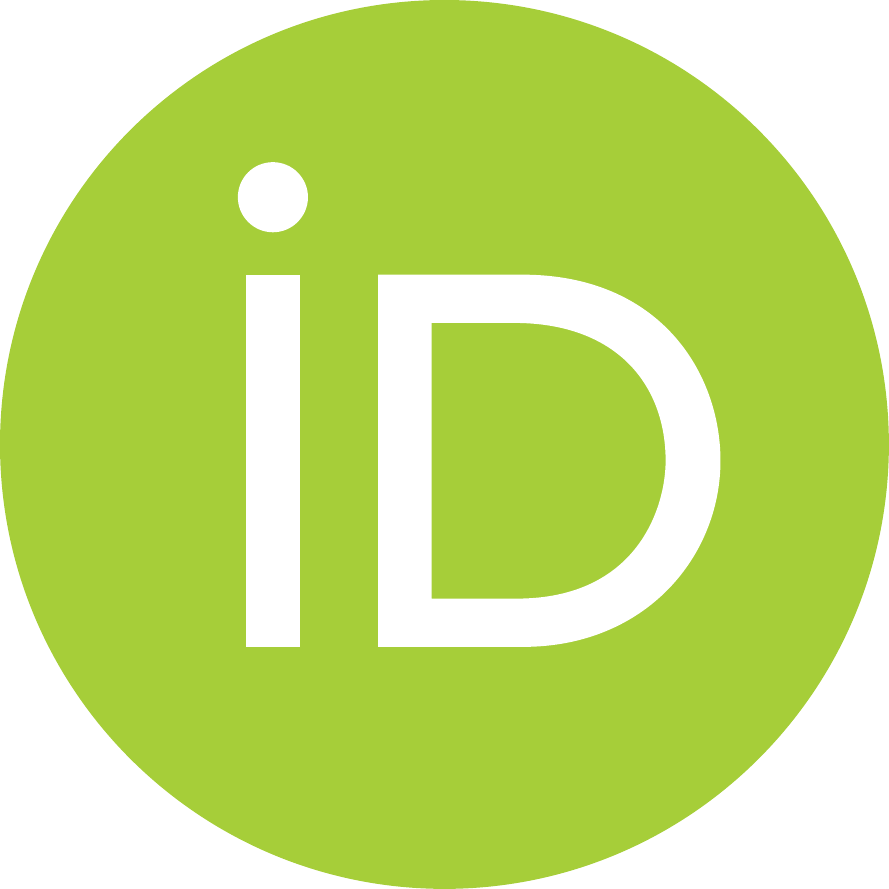}}}}

\newcommand\JWST{\emph{JWST}\xspace}
\newcommand\HST{\emph{HST}\xspace}
\newcommand\Spitzer{\emph{Spitzer}\xspace}

\title[Stellar populations from JWST]{JWST NIRCam+NIRSpec: Interstellar medium and stellar populations of young galaxies with rising star formation and evolving gas reservoirs}

\author[Tacchella et al.]{Sandro Tacchella\orcid{0000-0002-8224-4505},$^{1,2}$\thanks{E-mail: \href{mailto:st578@cam.ac.uk}{st578@cam.ac.uk}}
Benjamin D. Johnson\orcid{0000-0002-9280-7594},$^{3}$
Brant E. Robertson\orcid{0000-0002-4271-0364},$^{4}$
Stefano Carniani\orcid{0000-0002-6719-380X},$^{5}$
\newauthor
Francesco D'Eugenio\orcid{0000-0003-2388-8172},$^{1,2}$
Nimisha Kumari\orcid{0000-0002-5320-2568},$^{6}$
Roberto Maiolino\orcid{0000-0002-4985-3819},$^{1,2,7}$
Erica J. Nelson\orcid{0000-0002-7524-374X},$^{8}$
\newauthor
Katherine A. Suess\orcid{0000-0002-1714-1905},$^{9,10}$
Hannah \"Ubler\orcid{0000-0003-4891-0794},$^{1,2}$
Christina C. Williams\orcid{0000-0003-2919-7495},$^{11}$
Alabi Adebusola\orcid{0000-0002-3475-587X},$^{4}$
\newauthor
Stacey Alberts\orcid{0000-0002-8909-8782},$^{12}$
Santiago Arribas\orcid{0000-0001-7997-1640},$^{13}$
Rachana Bhatawdekar\orcid{0000-0003-0883-2226},$^{14}$
Nina Bonaventura\orcid{0000-0001-8470-7094},$^{15}$
\newauthor
Rebecca A. A. Bowler\orcid{0000-0003-3917-1678},$^{16}$
Andrew J. Bunker,$^{17}$
Alex J. Cameron\orcid{0000-0002-0450-7306},$^{17}$
Mirko Curti\orcid{0000-0002-2678-2560},$^{1,2}$
Eiichi Egami\orcid{0000-0003-1344-9475},$^{12}$
\newauthor
Daniel J. Eisenstein\orcid{0000-0002-2929-3121},$^{3}$
Brenda Frye\orcid{0000-0003-1625-8009},$^{12}$
Kevin Hainline\orcid{0000-0003-4565-8239},$^{12}$
Jakob M. Helton\orcid{0000-0003-4337-6211},$^{12}$
Zhiyuan Ji\orcid{0000-0001-7673-2257},$^{18}$
\newauthor
Tobias J. Looser,$^{1,2}$
Jianwei Lyu\orcid{0000-0002-6221-1829},$^{12}$
Michele Perna\orcid{0000-0002-0362-5941},$^{13}$
Timothy Rawle\orcid{0000-0002-7028-5588},$^{6}$
George Rieke\orcid{0000-0003-2303-6519},$^{12}$
\newauthor
Marcia Rieke\orcid{0000-0002-7893-6170},$^{12}$
Aayush Saxena\orcid{0000-0001-5333-9970},$^{7}$
Lester Sandles\orcid{0000-0001-9276-7062},$^{1,2}$
Irene Shivaei\orcid{0000-0003-4702-7561},$^{12}$
Charlotte Simmonds\orcid{0000-0003-4770-7516},$^{19}$
\newauthor
Fengwu Sun\orcid{0000-0002-4622-6617},$^{12}$
Christopher N. A. Willmer\orcid{0000-0001-9262-9997},$^{12}$
Chris J. Willott\orcid{0000-0002-4201-7367},$^{20}$ and
Joris Witstok\orcid{0000-0002-7595-121X}$^{1,2}$
\\
\\
\emph{\normalsize Affiliations are listed at the end of the paper}
}

\pubyear{2022}

\begin{document}
\label{firstpage}
\pagerange{\pageref{firstpage}--\pageref{lastpage}}
\maketitle

\begin{abstract}
We present an interstellar medium and stellar population analysis of three spectroscopically confirmed $z>7$ galaxies in the ERO \JWST/NIRCam and \JWST/NIRSpec data of the SMACS J0723.3-7327 cluster. We use the Bayesian spectral energy distribution (SED) fitting code \texttt{Prospector} with a flexible star-formation history (SFH), a variable dust attenuation law, and a self-consistent model of nebular emission (continuum and emission lines). Importantly, we self-consistently fit both the emission line fluxes from \JWST/NIRSpec and the broad-band photometry from \JWST/NIRCam, taking into account slit-loss effects. We find that these three $z=7.6-8.5$ galaxies ($M_{\star}\approx10^{8}~M_{\odot}$) are young with rising SFHs and mass-weighted ages of $3-4$ Myr, though we find indications for underlying older stellar populations. The inferred gas-phase metallicities broadly agree with the direct metallicity estimates from the auroral lines. The galaxy with the lowest gas-phase metallicity ($\mathrm{Z}_{\rm gas}=0.06~\mathrm{Z}_{\odot}$) has a steeply rising SFH, is very compact ($<0.2~\mathrm{kpc}$) and has a high star-formation rate surface density ($\Sigma_{\rm SFR}\approx22~\mathrm{M}_{\odot}~\mathrm{yr}^{-1}~\mathrm{kpc}^{-2}$), consistent with rapid gas accretion. The two other objects with higher gas-phase metallicity show more complex multi-component morphologies on kpc scales, indicating that their recent increase in star-formation rate is driven by mergers or internal, gravitational instabilities. We discuss effects of assuming different SFH priors or only fitting the photometric data. Our analysis highlights the strength and importance of combining \JWST imaging and spectroscopy for fully assessing the nature of galaxies at the earliest epochs. 
\end{abstract}

\begin{keywords}
early Universe -- galaxies: formation -- galaxies: evolution -- galaxies: high-redshift -- galaxies: star formation
\end{keywords}



\section{Introduction}

The \emph{James Webb Space Telescope} (\JWST) promises to reveal a new view of galaxy formation in the early Universe. For the first time, astronomers can now combine photometry (NIRCam) and spectroscopy (NIRSpec) in the rest-frame UV/optical to observe directly the growth of stellar populations and galactic structure in the first few hundred million years of cosmic history. The discoveries enabled by \JWST will allow for the formation history of stars, gas, metals, and dust to be detailed during the critical Epoch of Reionization \citep[EoR; for a review, see][]{robertson22}. Together, NIRCam and NIRSpec provide information on both broad-band stellar light and emission lines powered by Lyman Continuum (LyC) photons emitted by newly formed stars. This powerful combination allows us to precisely infer the properties of early galaxies without redshift uncertainties. The focus of this paper is to use jointly the initial \JWST Early Release Observations (EROs) NIRCam and NIRSpec data in the SMACS0723 cluster \citep{pontoppidan22} to infer the properties of galaxies with spectroscopic redshifts at $z\sim7.6-8.5$ and thereby gain a foothold on the physics of galaxy formation in the early Universe.

Detailed stellar population analyses of early galaxies advanced significantly during the pre-\JWST era thanks to combining \emph{Hubble Space Telescope} (\HST) and \emph{Spitzer Space Telescope} (\Spitzer) $3.6-4.5\mu$m data. Several studies used these datasets to constrain the stellar populations of $z>8$ galaxies, finding populations of young galaxies (mass-weighted stellar ages of a few tens of Myr; e.g., \citealt{stefanon21}), but also more evolved, older objects (with ages of a few hundred Myr; e.g., \citealt{hashimoto18, laporte21}). Significant challenges with these observations are the low signal-to-noise ratio, source blending as a result of the low spatial resolution of \Spitzer, and strong emission lines (i.e., H$\beta$ and [OIII]) that can mimic a strong Balmer/4000\AA\ break \citep{finkelstein13, labbe13, smit15, de-barros19, endsley21_ew}. Because of this, \citet{tacchella22_highz} and \citet{whitler23_sfh} have investigated how priors might influence the inference of the stellar ages and star-formation histories (SFHs) more generally, finding that the current \HST+ \Spitzer data are not very constraining in age-dating the highest-$z$ galaxies. Along these lines, the ultra-violet (UV) colours have been used to study the earliest phases of chemical enrichment \citep[e.g.,][]{wilkins11, finkelstein12a, bouwens14, bhatawdekar21}. However, the amount of attenuation, the attenuation law itself and the stellar and gas-phase metallicities remain largely unconstrained at $z>7$.

\JWST will improve the situation tremendously by delivering larger wavelength coverage and higher spectral resolution, both with spectra and medium/narrow band imaging. This will help break the degeneracy between strong emission lines and a strong Balmer/4000\AA\ break, which is needed to age-date these galaxies accurately \citep{roberts-borsani21_jwst, tacchella22_highz}, enabling not only with learning more about the physical properties of these high-$z$ galaxies during and before the EoR, but also with timing cosmic dawn by delivering accurate SFHs. Several \JWST programmes will deliver such datasets (i.e. spectra and medium band imaging), including the \textit{JWST Advanced Deep Extragalactic Survey} (JADES), which is a Guaranteed Time Observer (GTO) programme using about 800 hours of prime time and 800 hours of parallel time to study the formation and evolution of galaxies, combining NIRSpec, NIRCam, and MIRI data in a coordinated observing programme \citep{rieke19}. 

In this work, we make use of the \JWST ERO of the SMACS J0723.3-7327 cluster field \citep{pontoppidan22}. The ERO multi-mode observations of SMACS J0723.3-7327 include NIRCam and MIRI imaging, NIRSpec Multi-Object Spectroscopy, and NIRISS Wide-Field Slitless Spectroscopy of the cluster and surrounding field. The NIRSpec MSA configuration was designed based on a photometric catalogue constructed from the NIRCam imaging, following a workflow similar to that of future science programs such as JADES. We focus on the three galaxies that have a spectroscopic redshift $z>7$, namely Galaxy ID 04590, 06355 and 10612 (Table~\ref{tab:galaxies}).

These three objects have been discussed in several recent papers \citep[e.g.,][]{arellano-cordova22, brinchmann22, carnall23_ero, curti23, katz23, rhoads23, schaerer22, trump23, trussler22}. The different studies rely on different measurement techniques, but also on different reduction and calibration pipelines, making comparisons difficult. \citet{carnall23_ero} performed SED fitting of the NIRCam imaging and the existing \HST imaging for these three objects, finding low stellar masses ($\log(M_{\star}/M_{\odot})=7.1-8.2$) and correspondingly young mean stellar ages of only a few Myr. \citet{curti23} used the [OIII]4363 auroral line to measure metallicities with the direct $T_{\rm e}$ method, finding Galaxy ID 04590 to be extremely metal poor ($12+\log(\mathrm{O}/\mathrm{H})\approx7$), Galaxy ID 10612 to be about 1/10 solar and Galaxy ID 06355 to be about one-third solar. The latter two objects are marginally consistent with the Fundamental Metallicity Relation \citep[FMR;][]{mannucci10, onodera16, curti20, suzuki21, papovich22_clear}, while Galaxy ID 04590 deviates significantly from it, consistent with being far from the smooth equilibrium between gas flows, star formation and metal enrichment in place at later epochs. This possibly indicates that we enter at $z>7$ another epoch of galaxy formation, where galactic systems are out of equilibrium and evolve more stochastically. Consistent with this, other studies find high ionization, low metallicity and high pressure in the interstellar medium \citep[ISM;][]{katz23, schaerer22, rhoads23, trump23, trussler22}. 

In this paper, we contribute to these interesting results by performing a full spectro-photometric analysis by simultaneously fitting the NIRCam photometry and the NIRSpec spectroscopy, taking into account slit-loss effects. We confront the measured gas-phase metallicity with the inferred SFHs and morphology of these three $z=7.6-8.5$ galaxies, highlighting the strength and importance of combining \JWST imaging and spectroscopy for fully assessing the nature of galaxies at the earliest epochs. Specifically, Section~\ref{sec:data} presents the NIRCam and NIRSpec data processing and describes the details of the stellar population inference from the spectro-photometric modelling. We present our key results in Section~\ref{sec:results} and conclude in Section~\ref{sec:conclusions}. We assume the latest {\it Planck} flat $\Lambda$CDM cosmology with $\mathrm{H}_{0}=67.36$, $\Omega_m=0.3153$, and $\Omega_{\Lambda}=0.6847$ \citep{planck-collaboration20} and $\mathrm{Z}_{\odot}=0.0142$. All quantities are corrected for magnification if not otherwise indicated (see Section~\ref{subsec:magnification}). When quoting values of derived quantities, we quote the median and $16-84\%$ quantile-based errors if not otherwise stated.

\section{Data and Method}
\label{sec:data}

\begin{figure*}
	\includegraphics[width=\textwidth]{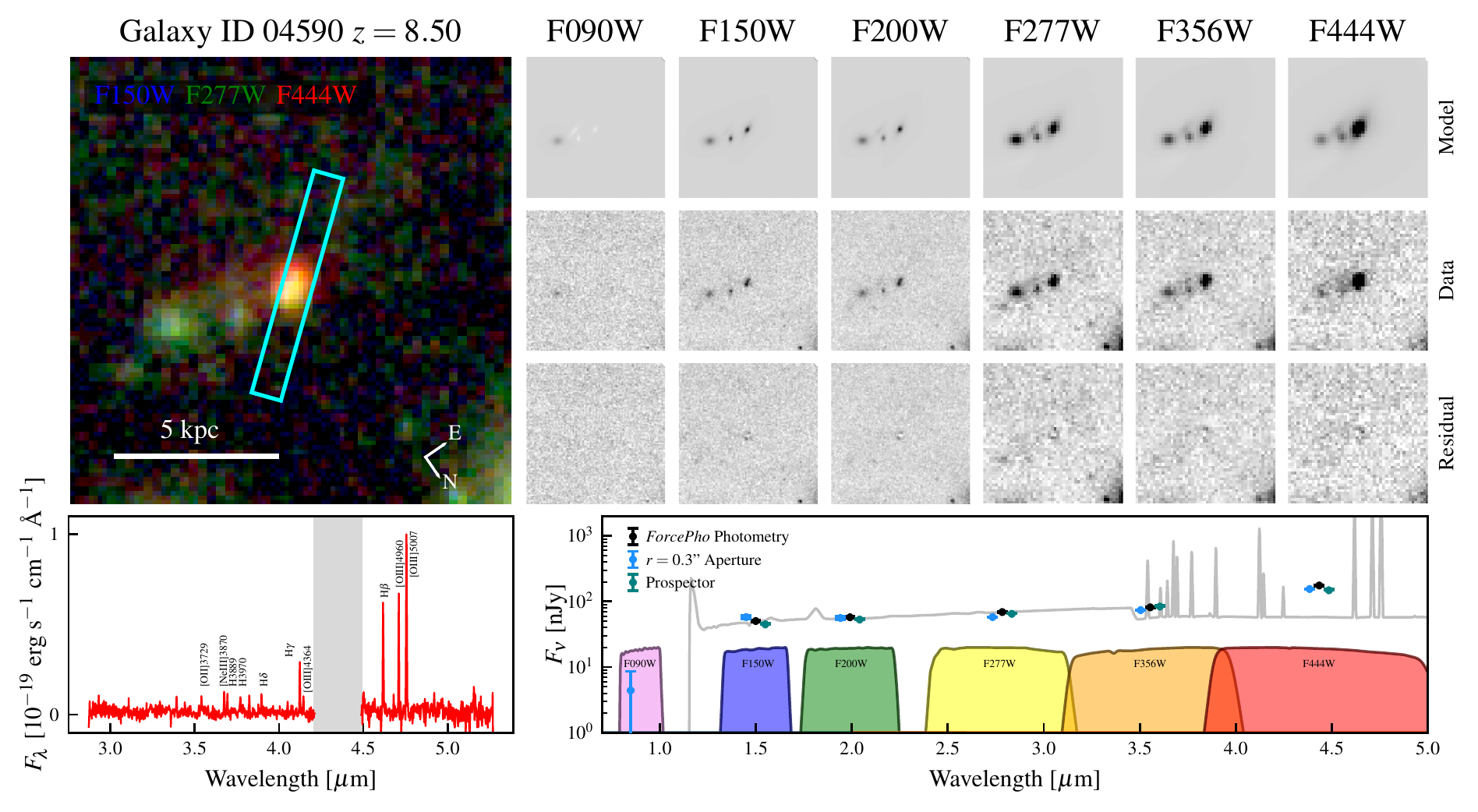}
	\caption{Composite imaging with NIRSpec slit and photometric data for Galaxy ID 04590 at redshift $z=8.50$. Galaxy ID 04590 is the centre of the F090W-F150W-F444W false-colour image on top of which the NIRSpec slit falls (upper left; scale bar and compass provided for reference). The source on the very left is a lower-redshift object given that it is visible in the F090W filter (see images on the upper right, middle row), while the source directly on the left might be at a similar redshift, but is not considered part of Galaxy ID 04590. The small residuals between the model and the data demonstrate the effectiveness of \texttt{forcepho} in capturing the multiband scene. \texttt{forcepho} may also be used to model the source photometry with high-accuracy (bottom right; black points correspond to the top component of the \texttt{forcepho} model. NIRCam filter transmissivity curves are shown as shaded regions) and is in rough agreement with 0.3'' radius aperture photometry. The \texttt{forcepho} photometry is used in conjunction with the NIRSpec spectrum (lower left; red line. Shaded region shows the chip gap.) to supply the empirical constraints on our \texttt{Prospector} spectral energy distribution model (lower right; gray line). Both the photometry and emission line strengths are fit simultaneously by \texttt{Prospector}. The resulting SED model reveals a star-forming galaxy with strong emission line flux contributions to the rest-frame optical light.}
    \label{fig:data4590}
\end{figure*}

\begin{figure*}
	\includegraphics[width=\textwidth]{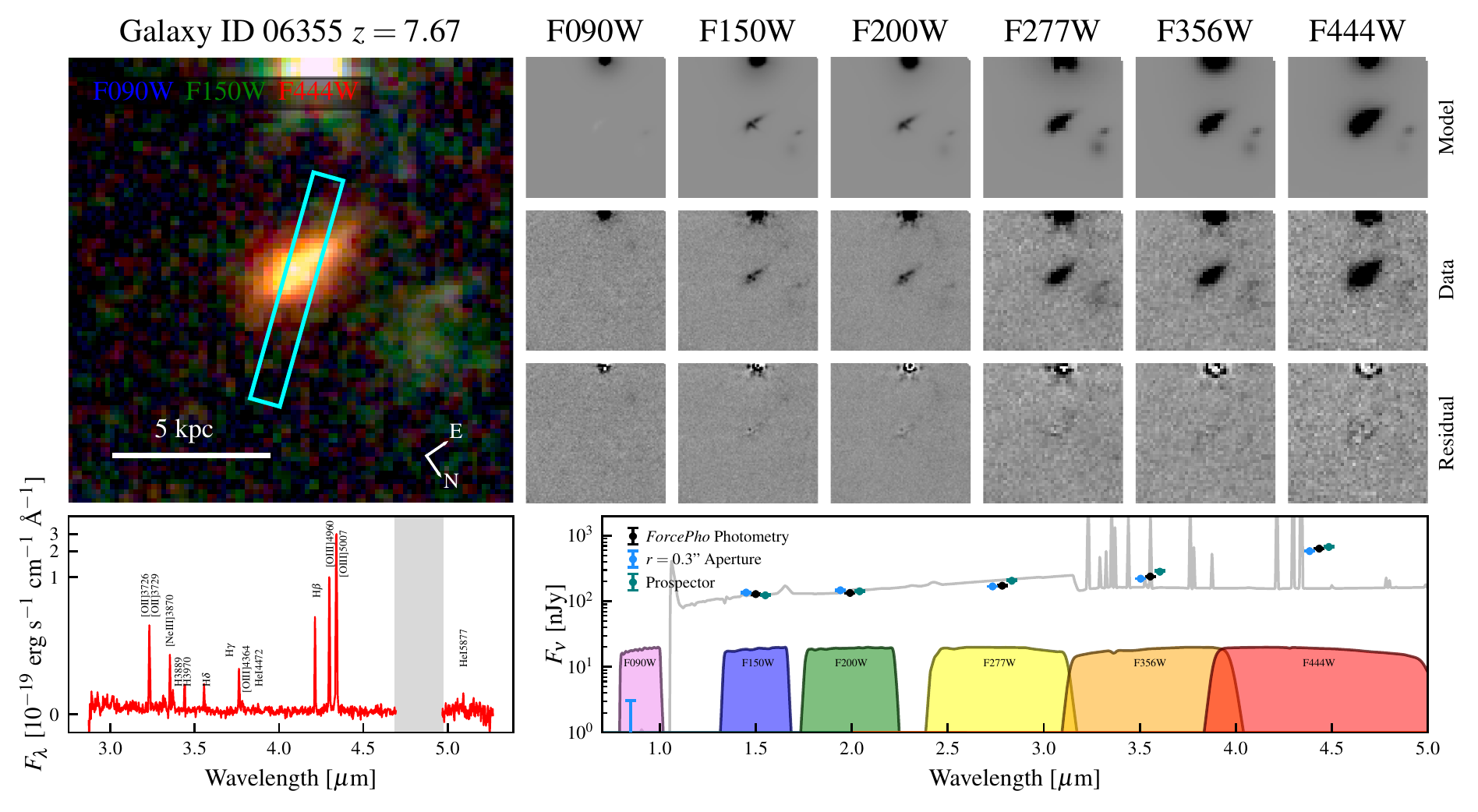}
	\caption{Same as Fig.~\ref{fig:data4590}, but for Galaxy ID 06355 at redshift $z=7.67$. This galaxy is clearly more extended than Galaxy ID 04590 and consistent of multiple components.}
    \label{fig:data6355}
\end{figure*}

\begin{figure*}
	\includegraphics[width=\textwidth]{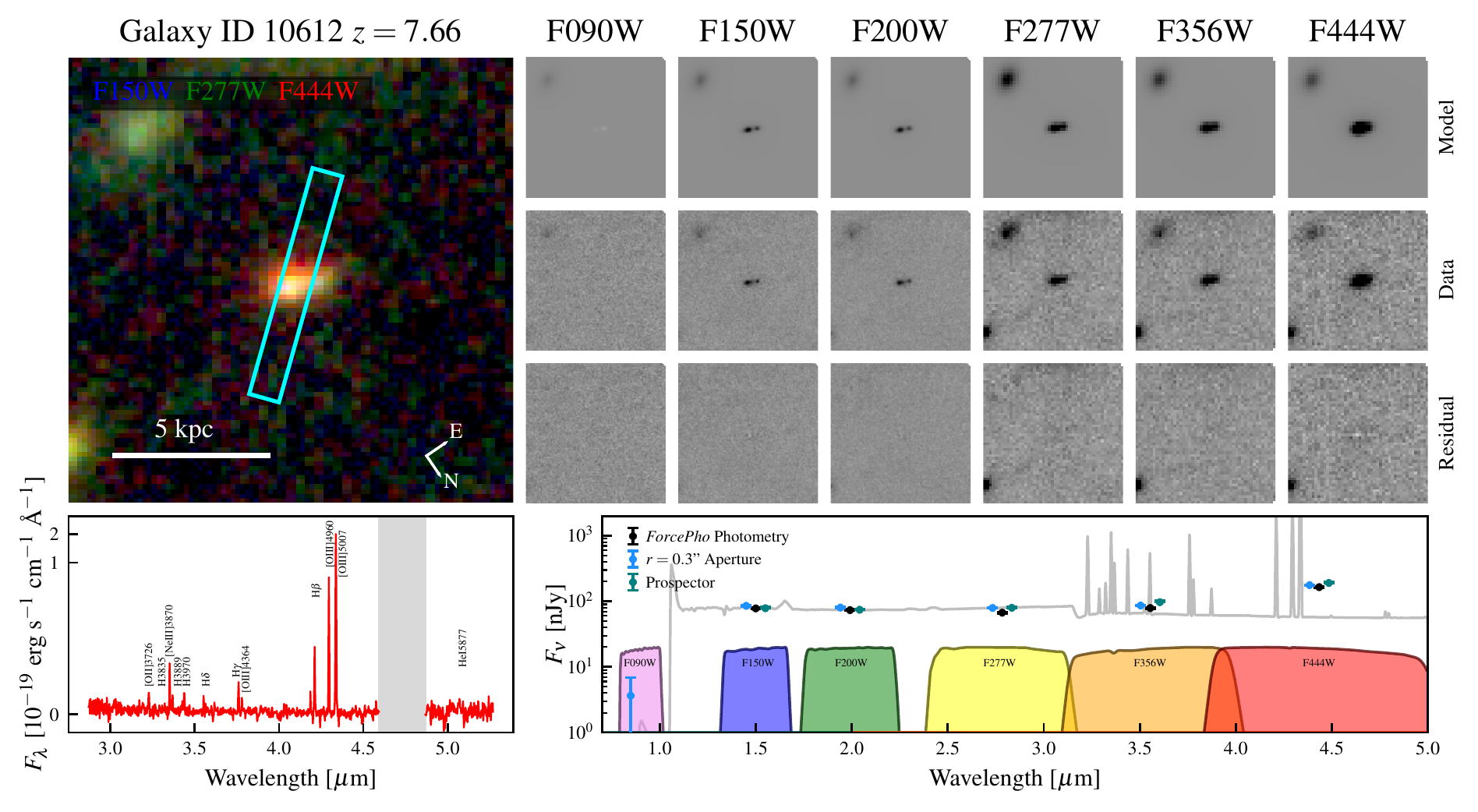}
	\caption{Same as Fig.~\ref{fig:data4590}, but for Galaxy ID 10612 at redshift $z=7.66$. This galaxy is consists of two components.}
    \label{fig:data10612}
\end{figure*}

All \JWST observations used in this work are taken as part of the SMACS0723 ERO (Programme \#2736; \citealt{pontoppidan22}). We focus in this analysis on three galaxies that are both covered with NIRCam and NIRSpec and lie at $z>7$ (Tab.~\ref{tab:galaxies}).

\subsection{NIRCam imaging and photometry}
\label{subsec:nircam}

We use the deep NIRCam imaging data of SMACS0723 in the F090W, F150W, F200W, F277W, F356W, and F444W filters, which cover an observed wavelength range of $\lambda_{\rm obs}=0.8-5\upmu\mathrm{m}$. The images reach (and surpass) $AB\approx29$ mag. We reduce the raw level-1 data products with the public \JWST pipeline (v1.9.2; \url{https://github.com/spacetelescope/jwst}), using the latest available calibration files (CRDS\_CTX=jwst\_1039.pmap). An additional, local background subtraction is performed on the final mosaiced images and applied to the individual exposures, see details in \citet{robertson23} and \citet{tacchella23}.

We measure object fluxes using the Bayesian model-fitting program \texttt{forcepho} (Johnson et al., in prep.).  This program simultaneously fits multiple PSF-convolved S\'{e}rsic profiles, each having 11 parameters, to all of the background-subtracted stage 2 \texttt{cal} exposures for every NIRCam filter. The joint posterior distribution for the parameters of all of the profiles is sampled via Markov Chain Monte Carlo (MCMC), allowing estimation of covariances between the parameters of a single object (such as S\'{e}rsic index and half-light radius) as well as covariance between objects (due to blending), and identification of any non-Gaussian posterior distributions. The simultaneous fit to all bands and the MCMC sampling allow us to obtain self-consistent estimates of object fluxes in each band. We represent Galaxy ID 06355 and 10612 with several sub-components to account for clumpy structure obvious in the higher resolution F150W imaging, and we simultaneously model all sources within 1\arcsec\ to account for the contribution of extended structures. To determine total source fluxes we then combine the fluxes of the sub-components while taking into account covariances among them, which we accomplish by summing the sub-component fluxes for each posterior sample and then computing the mean and standard deviation of the summed fluxes. 
In Figures \ref{fig:data4590}, \ref{fig:data6355}, and \ref{fig:data10612} we show examples of the data, model, and stacked residuals for one draw from the MCMC chain.  We have varied the number of sub-components used for the sources and find similar total fluxes. Due to uncertainties in photometric calibration and other factors affecting these very early data, we adopt an error floor of 10\% on the photometry.

The \texttt{forcepho} fits to the images also provide size estimates.  For the single component of Galaxy ID 04590 this is taken directly from the half-light radius of the fitted S\'{e}rsic profile.  For the other two Galaxy IDs that have multiple components, we estimate a combined effective half-light radius by generating a model image with the PSF removed using the F150W fluxes, compute the barycenter, and determine the radius containing half the combined model flux.  This yields half-light radii of $0.039''\pm0.004$ ($0.18\pm 0.02$ kpc), $0.077''\pm 0.008$ ($0.39\pm 0.04$ kpc) and $0.100''\pm 0.010$ ($0.51\pm 0.05$ kpc) respectively for Galaxy ID 04590, 06355, and 10612. The size for Galaxy ID 10612 is driven by the separation between the two dominant components, which are each $\lesssim 0.03''$ in half-light radius. We note that we have not corrected these sizes for lensing distortions (Section~\ref{subsec:magnification}). We expect this effect to be small for Galaxy ID 06355 and 10612, because they are only weakly magnified. This effect could however be more substantial for Galaxy ID 04590, indicating that our size is possibly an upper limit.

\subsection{NIRSpec spectroscopy}
\label{subsec:nirspec}

NIRSpec observations for SMACS0723 were carried out using two disperser/filter combinations, but given the redshift of our targets, in this work we use only data from the G395/F290LP combination, covering the wavelength range 2.9--5.2~$\upmu$m with a spectral resolution $R\sim1000$. The observations consist of two pointings, with a total integration of 8,840 seconds. We show the slit position in In Figures \ref{fig:data4590}, \ref{fig:data6355}, and \ref{fig:data10612}. For all three galaxies, the slits cover the main part of the galaxies, which supports our assumption of modelling the photometry together with the spectroscopy (i.e., treating the spectroscopic observations as representative of the galaxy). Nevertheless, slit losses are important (see below). We estimate the slit loss to be 12\%, 29\% and 39\% for Galaxy ID 4590, 6355, and 10612, respectively, using the F356W band, which probes the wavelength range of the spectrum.

In this paper, we use the fully reduced spectra from \citet{curti23}\footnote{The 1-d spectra are publicly available at \url{https://doi.org/10.5281/zenodo.6940561}}, which are based on level-2 data from the MAST archive and were processed using the GTO  pipeline \citetext{NIRSpec/GTO collaboration, in prep.; \citealp{ferruit22}}. The data and data reduction procedure are already described in \citet{curti23}, so here we report only a summary.

The GTO pipeline uses a custom flagging and masking algorithm for bad pixels and cosmic rays, which effectively removes all artefacts from the final spectra. The extraction is performed with a custom aperture, using a boxcar extraction. After inspecting the exposures of the individual nods, it was found that one of the shutters on which Galaxy ID 04590 was nodded did not open in Obs~7 \citep{rawle18}; the affected nod positions were discarded before stacking the data. Stacking was performed with the GTO pipeline, taking into account the variance and quality arrays of both pointings. We refined the flux calibration using an empirical response function, derived from a spectro-photometric star observed during commissioning \citep[Programme \#1128,][]{gordon22}. Because our three targets are only marginally resolved, we apply the path-loss correction derived for a point-like source. The path-loss correction appropriate for an extended uniform source gives a spectrum that has higher overall flux, but approximately the same shape as our default choice\footnote{This is because the wavelength range covered by our spectra is sufficiently narrow that wavelength-dependent path losses do not affect significantly our results.}. As we discuss below (see Section~\ref{subsec:prospector}), we introduce a nuisance parameter in the spectro-photometric modelling to account for any mismatch between the photometry and spectroscopy (related to slit loss, calibration issues as well as physical effects). Therefore, it does not matter in this analysis whether we use the point-like or extended source approximation. The resulting spectra are shown in the lower-left corner of Figures~\ref{fig:data4590}--\ref{fig:data10612}.

Emission-line fluxes were measured using \textsc{ppxf} \citep{cappellari17}, using the configuration described in \citet{curti23}. Briefly, we fit the stellar continuum using a linear combination of simple stellar-population spectra from the C3K library \citep{conroy19}, using the MIST isochrones \citep{choi16}. The continuum is scaled using a 10\textsuperscript{th}-order Legendre polynomial. Emission-lines are modelled as Gaussians. All the model spectra (both continuum and emission-line) are constrained to have the same velocity and velocity dispersion; note that the continuum is only marginally resolved; replacing the SSP template spectra with a constant does not change the emission-line fluxes \citep{curti23}. The resulting fluxes have already been published \citep[][their Table~1]{curti23} and are not repeated here. We remark that fitting each line individually does not change our conclusions, as reported in \citet{katz23}. The spectroscopic redshifts in this work are based on the emission-line velocities measured by \textsc{ppxf}.

In the fitting, we mask the following emission lines: [OIII]4364, H$\delta$, H$\zeta$, and [NeIII]3870. We mask [OIII]4364 because we cannot reproduce very high [OIII]4364 emission line fluxes (as measured here) with our current \texttt{cloudy}-based modelling for the nebular emission (see next section). As discussed in \citet{dors11}, photoionization models have trouble predicting [OIII]4364 at a necessary accuracy (see also \citealt{brinchmann22}, who has also excluded this line from their modelling). Directly related, \citet{katz23} show that in \texttt{cloudy} one can add an extra, hard ionization source, which then, for fixed ionization parameter and gas-phase metallicity, increases [OIII]4363/[OIII]5007 without really changing the other line ratios, notably [OIII]5007/H$\beta$ or [OIII]5007/[OII]3727, which are basically the only other metallicity sensitive lines we have in our spectrum (at least at the metallicities that are relevant). So we conclude that even when masking [OIII]4363, we should recover reliable gas-phase metallicity estimates.

Furthermore, we mask H$\delta$ because this line is abnormally weak, i.e. about 25\% weaker than what it should be given the observed Balmer decrement on H$\gamma$. As discussed in \citet{curti23}, this might be because of possible background subtraction issues. Finally, we mask both H$\zeta$ and [NeIII]3870 because they are blended and the current decomposition method seems to be unable to provide reliable fluxes for each of the lines (see also Section~\ref{subsec:runs}).

\begin{figure*}
	\includegraphics[width=\textwidth]{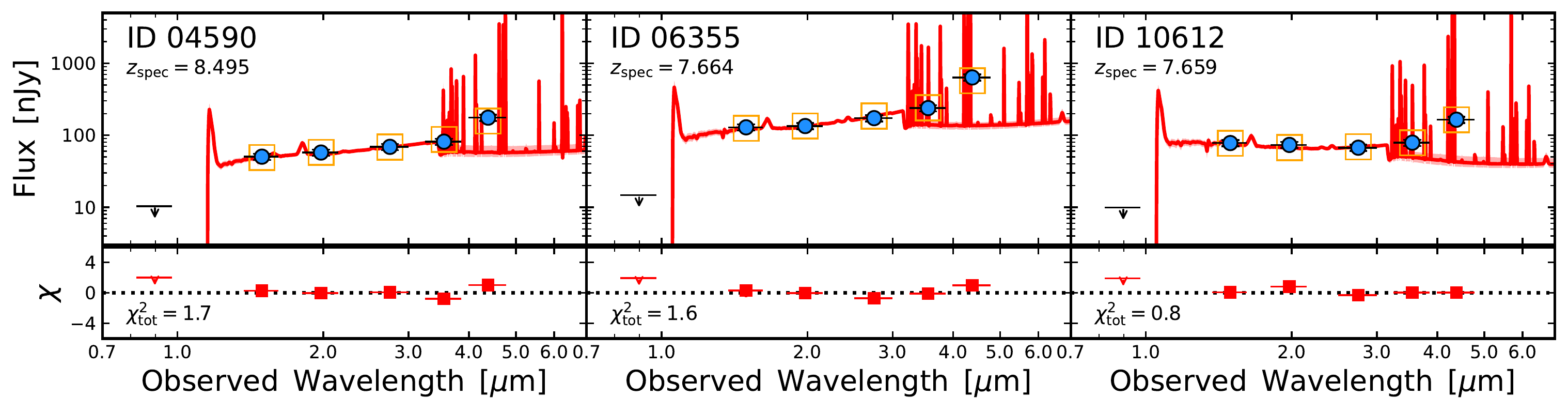}
	\includegraphics[width=\textwidth]{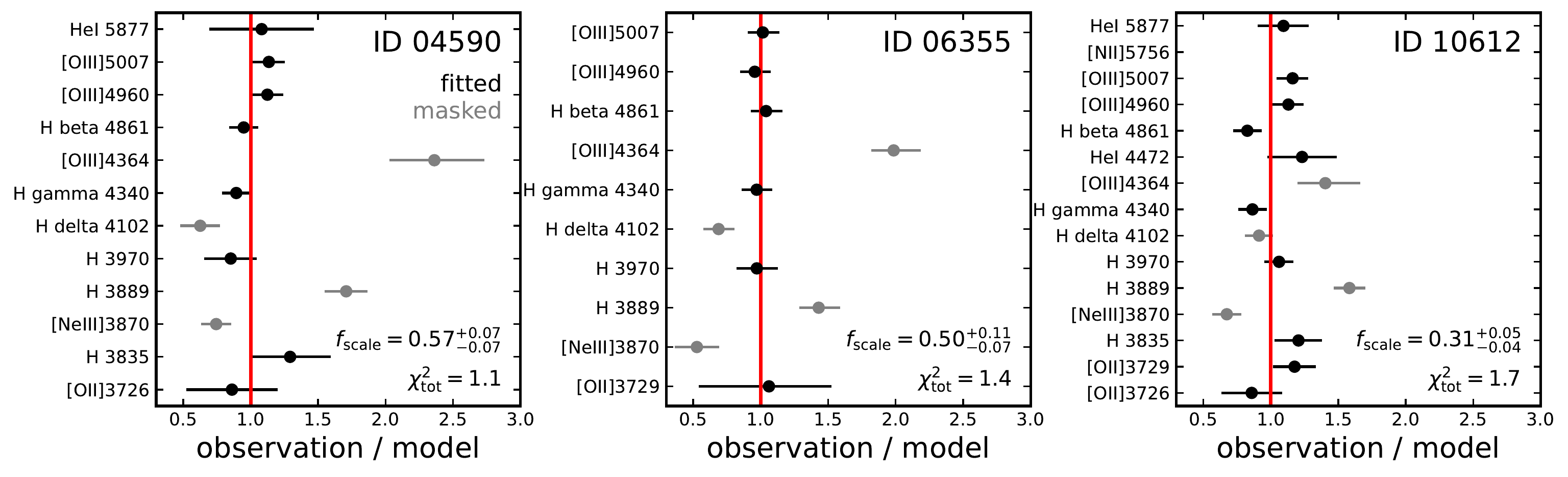}
    \caption{Comparison of inferred model with the observed data. \textit{Top panels:} SED of the galaxies. Blue points are the observed photometry from NIRCam, while the orange boxes mark the posterior photometry from our model. The red solid line (and the shaded region) is the median SED of the posterior (and the 16th-84th percentile). The $\chi^2$ values indicate that our fits are reasonable. \textit{Bottom panels:} NIRSpec observations versus model posteriors for the emission lines. The black circles mark the lines that are fitted, while the grey points indicate the masked emission lines, which have not been fitted. Overall, the emission lines are well reproduced with our model after rescaling the model-based fluxes by $f_{\rm scale}$, which can be motivated by slit-losses and physical effects such as LyC dust absorption or LyC escape.}
    \label{fig:sed}
\end{figure*}

\begin{table*}
	\centering
	\caption{Properties of the three galaxies of this work. The columns are the ID, redshift, magnification ($\mu$), right ascension (RA), declination (DEC), stellar mass, SFR averaged over 10 Myr, stellar age (half-mass time $t_{50}$), dust attenuation in the V-band (A$_{\rm V}$), gas-phase metallicity ($\mathrm{Z}_{\rm gas}$), stellar metallicity ($\mathrm{Z}_{\star}$), and effective size ($R_{\rm e}$). All quantities here are corrected for magnification.}
	\label{tab:galaxies}
	\begin{tabular}{ccccccccccccc}
		\hline
		ID & Redshift & $\mu$ & RA & DEC & $\log(M_{\star})$ & SFR$_{10}$ & $t_{50}$ & A$_{\rm V}$ & $\log(\mathrm{Z}_{\rm gas})$ & $\log(\mathrm{Z}_{\star})$ & $R_{\rm e}$ \\
		  &  &  & [deg] & [deg] & [$M_{\odot}$] & [$\mathrm{M}_{\odot}/\mathrm{yr}$] & [Myr] & [mag] & [$\mathrm{Z}_{\odot}$] & [$\mathrm{Z}_{\odot}$] & [kpc] \\
		\hline
		04590 & 8.4953 & $3.74\pm0.07$ & 110.85933 & -73.44916 & $7.8^{+0.4}_{-0.2}$ & $5_{-1}^{+2}$ & $4_{-3}^{+177}$ & $0.61_{-0.19}^{+0.22}$ & $-1.2_{-0.1}^{+0.1}$ & $-1.2_{-0.3}^{+0.3}$ & $0.18\pm 0.02$ \\
		06355 & 7.6643 & $1.23\pm0.01$ & 110.84452 & -73.43508 & $8.6^{+0.2}_{-0.2}$ & $31_{-6}^{+12}$ & $3_{-1}^{+186}$ & $0.43_{-0.15}^{+0.26}$ & $-0.6_{-0.1}^{+0.1}$ & $-0.8_{-0.2}^{+0.2}$ & $0.39\pm 0.04$ \\
		10612 & 7.6592 & $1.34\pm0.01$ & 110.83395 & -73.43454 & $7.8^{+0.5}_{-0.1}$ & $5_{-1}^{+1}$ & $3_{-1}^{+59}$ & $0.08_{-0.05}^{+0.08}$ & $-0.6_{-0.1}^{+0.1}$ & $-0.7_{-0.4}^{+0.2}$ & $0.51\pm 0.05$ \\
		\hline
	\end{tabular}
\end{table*}

\begin{table*}
	\centering
	\caption{Effect on the stellar populations (stellar mass, SFR, and stellar age) when including/excluding emission lines (EL versus photometry-only) and changing the SFH prior (bursty versus continuous). All quantities here are corrected for magnification.}
	\label{tab:model_variation}
    \begin{tabular}{|l|c|c|c|c|c|c|c|c|c|c|c|c|}
    \hline
    \multicolumn{1}{|c|}{} & \multicolumn{3}{c|}{EL \& bursty} & \multicolumn{3}{c|}{EL \& continuous} & \multicolumn{3}{c|}{phot-only \& bursty} & \multicolumn{3}{c|}{phot-only \& continuous}\\
    \cline{2-4} \cline{5-7} \cline{8-10} \cline{11-13}
    \multicolumn{1}{|c|}{ID} & $\log(M_{\star})$ & SFR$_{10}$ & $t_{50}$ & $\log(M_{\star})$ & SFR$_{10}$ & $t_{50}$ & $\log(M_{\star})$ & SFR$_{10}$ & $t_{50}$ & $\log(M_{\star})$ & SFR$_{10}$ & $t_{50}$ \\
    \hline
    04590 & $7.8^{+0.4}_{-0.2}$ & $5_{-1}^{+2}$ & $4_{-3}^{+177}$ & $8.4_{-0.6}^{+0.4}$ & $5_{-1}^{+3}$ & $95_{-88}^{+103}$ & $8.2_{-0.6}^{+0.9}$ & $3_{-3}^{+2}$ & $42_{-39}^{+175}$ & $8.9_{-0.6}^{+0.3}$ & $3_{-2}^{+3}$ & $177_{-100}^{+65}$ \\
    06355 & $8.6^{+0.2}_{-0.2}$ & $31_{-6}^{+12}$ & $3_{-1}^{+186}$ & $8.7_{-0.2}^{+0.6}$ & $30_{-4}^{+11}$ & $4_{-1}^{+150}$ & $8.9_{-0.4}^{+0.4}$ & $36_{-9}^{+15}$ & $14_{-11}^{+295}$ & $8.9_{-0.4}^{+0.4}$ & $40_{-12}^{+21}$ & $29_{-26}^{+133}$ \\
    10612 & $7.8^{+0.5}_{-0.1}$ & $5_{-1}^{+1}$ & $3_{-1}^{+59}$ & $8.4_{-0.2}^{+0.2}$ & $5_{-1}^{+1}$ & $126_{-90}^{+102}$ & $7.9_{-0.2}^{+0.5}$ & $5_{-1}^{+1}$ & $4_{-1}^{+116}$ & $8.5_{-0.4}^{+0.3}$ & $6_{-1}^{+2}$ & $143_{-133}^{+95}$
 \\
    \hline
    \end{tabular}
\end{table*}

\subsection{Modelling of the spectro-photometric data}
\label{subsec:prospector}

We use the SED fitting code \texttt{Prospector} \citep{johnson21} to  simultaneously fit the photometric and spectroscopic data. We adopt a similar, 13-parameter physical model and priors as described in \citet{tacchella22_highz}. Specifically, we fit for the stellar mass, stellar and gas-phase metallicities, dust attenuation (using a two-component dust model, which includes a diffuse component for the entire galaxy, extra attenuation around the youngest stars ($<10~\mathrm{Myr}$), and a flexible slope; 3 free parameter) and ionization parameter for the nebular emission. We adopt the flexible SFH prescription \citep{leja19_nonparm} with 6 time bins with the bursty-continuity prior \citep{tacchella22_highz}. This model includes 5 free parameters which control the ratio of SFR in six adjacent time bins; the first two bins are spaced at $0-5~\mathrm{Myr}$ and $5-10~\mathrm{Myr}$ of lookback time, and the remaining four bins are log-spaced to $z=20$. This implies that we are not able to infer ages below 2.5 Myr, which we believe is a realistic lower limit for galaxies. We also fit for a  nuisance parameter that scales all the emission line fluxes by a constant factor to account for flux calibration offsets or, to first order, slit losses or physical effects (see below for details). For all the fits, we assume the MESA Isochrones and Stellar Tracks (MIST) stellar models \citep{choi17} and a \citet{chabrier03} initial mass function (IMF).

\texttt{Prospector} has the ability to model (and fit) nebular emission line fluxes using the nebular line and continuum emission predictions provided by the Flexible Stellar Population Synthesis (FSPS) code, as described in \citet{byler17}.  These predictions are based on interpolated \texttt{cloudy} \citep[version 13.03;][]{ferland13} photo-ionization models that were constructed on a grid of ionization parameter ($U$), gas-phase metallicity, and Single Stellar Population (SSP) age, using stellar spectra of the same age and metallicity as the ionizing source. The solar nebular abundances are taken from \citet{anders89}. The \texttt{cloudy} modelling used a constant electron density and the highest ionization parameter in the grid is $\log(U)=-1$, making it difficult to model extreme [OIII] ratios \citep[e.g.][]{ferland13, katz23}. Critically, the nebular emission is scaled by the number of ionizing photons predicted at each SSP age for the given SFH.

Crucial assumptions in this model for the emission lines are that all emission lines are powered by star formation, the LyC escape fraction is zero, and there is no dust absorption of LyC. Under these assumptions the predicted nebular emission line fluxes are consistent with the ionizing flux of the modelled stellar population. As discussed in \citet[][see also \citealt{smith22}]{tacchella22_Halpha}, these assumptions do not necessarily hold at these early epochs: LyC absorption by dust can be important (absorbing up to 50\% of the LyC photons during a starburst phase), about $5-10\%$ of the LyC photons are absorbed by helium, and a small fraction of the LyC can also escape the galaxy. However, we add additional flexibility in two ways. First, we allow the gas-phase and stellar metallicity to take on different values, which in detail means that the shape of the model ionizing spectrum may be different than that used to predict the nebular lines ratios, though the normalisation (and hence Balmer line flux) is kept consistent.  Second, we rescale all the model emission-line fluxes multiplying them by $\text{f}_\mathrm{scale}$, a fittable nuisance parameter (we do not rescale the nebular continuum).  This constant factor can account for flux calibration offsets with NIRSpec or, to first order, slit losses, differential magnification effects\footnote{Our sources are extended, consisting of several components. Different emission lines have probably different spatial extents \citep[][their Figure~6]{perez-montero07}. Since the magnification might not be constant across the source, emission lines from more compact regions could be more magnified, leading to a bias in the line ratios (``magnification bias''). This effect is probably negligible for Galaxy ID 06355 and ID 10612, which both have a very low magnification (Table~\ref{tab:galaxies}).} or physical effects like LyC escape or dust absorption.

\subsection{Different runs}
\label{subsec:runs}

In addition to the fiducial \texttt{Prospector} setup described above, we run three further configurations in order to investigate how our results depend on the modelling assumptions as well as the data present. Specifically, our fiducial approach simultaneously fits the NIRCam photometry and NIRSpec emission lines with the bursty-continuity prior (``EL \& bursty'' fits). Since the inferred stellar population parameters can significantly depend on the assumed SFH prior \citep{carnall19_sfh, leja19_nonparm, suess22_prior, tacchella22_quench, tacchella22_highz, whitler23_sfh}, we vary the SFH prior. Specifically, the bursty-continuity SFH prior assumes that $\Delta\log(\mathrm{SFR})$ between adjacent time bins is described by Student's t-distribution with $\sigma=1.0$ and $\nu=2$. We also run fits with the standard continuity prior (``EL \& continuous'' fits), which assumes $\sigma=0.3$, which weights against sharp transitions between the different time bins of the SFH \citep{leja19_nonparm}. In addition, to study the effect of including emission lines in the fitting, we fit those two SFH prescriptions to only the photometry, i.e. ignoring the emission line constraints (``phot-only \& bursty'' and ``phot-only \& continuous''). 

The key results of our fiducial fits are tabulated in Table~\ref{tab:galaxies}, while Table~\ref{tab:model_variation} shows the effects on the inferred stellar masses, SFRs and stellar ages when varying the SFH prior and including/excluding emission lines. We present and discuss these results in Section~\ref{sec:results}. We focus on the impact of including emission lines in the fitting in Sections~\ref{subsec:EL} and discuss the SFH prior in Section~\ref{subsec:sfh_prior}.

Figure~\ref{fig:sed} shows the goodness of our fiducial fits (i.e. fitting both photometry and emission lines with the bursty SFH prior): the top panels compare the inferred model and observed SEDs, while the bottom panels compare the individual emission lines. In the top panels, the observed NIRCam photometry obtained with \texttt{forcepho} is indicated with blue circles, while the inferred model photometry is shown as orange squares. The absolute values of $\chi$ for the individual bands are typically $<1$ and the total $\chi^2_{\rm tot}$ values are 1.7, 1.6 and 0.8 for Galaxy ID 04590, 06355 and 10612, respectively.   

The emission lines are overall well reproduced (bottom panels of Figure~\ref{fig:sed}) with $\chi^2_{\rm tot}=1.1$, 1.4 and 1.7, respectively. As mentioned in Section~\ref{subsec:nirspec}, we have masked [OIII]4364, H$\delta$, H$\zeta$, and [NeIII]3870 (plotted as grey points). Our inferred model indeed underpredicts the emission line [OIII]4364 by $50-250\%$, while H$\delta$ is underpredicted by $10-30\%$. Furthermore, the deblended lines H$\zeta$ and [NeIII]3870 are under- and over-predicted, respectively. 

Importantly, the bottom panels of Figure~\ref{fig:sed} include the nuisance parameter that rescales all the emission fluxes. We find that the emission lines predicted from the model need to be suppressed by $40-70\%$ in order to be consistent with the observed fluxes. Not including this nuisance parameter would lead to nonphysical high stellar metallicities (i.e. $>\mathrm{Z}_{\odot}$) in order to suppress the ionizing photon production. Specifically, the rescale factors $f_{\rm scale}$ are $0.57_{-0.07}^{+0.07}$, $0.50_{-0.07}^{+0.11}$, and $0.31_{-0.04}^{+0.05}$ for Galaxy ID 04590, 06355 and 10612, respectively. Our inferred rescale factors are physically plausible. They are larger than the slit loss corrections (12\%, 29\% and 39\% for Galaxy ID 4590, 6355, and 10612), implying that the slit loss can only partially explain our inferred $f_{\rm scale}$, and that the remaining 31\%, 21\% and 30\% need be explained via other means, for example Lyman continuum absorption by dust or escape. Overall, this highlights that understanding slit loss corrections is fundamentally important to shed light onto both the stellar populations but also radiative transfer effects.

Finally, we compare the goodness of the resulting fits with the four different setups introduced above. The current uncertainties and this small sample do not allow us to infer whether any of the models is preferred (i.e. Bayes factor is inconclusive). However, we find that the bursty SFH prior consistently produces comparable or lower $\chi^2$ values both when including or excluding the emission lines in the fitting than the standard continuity prior: we find -- for Galaxy ID 04590, 06355 and 10612 -- $\chi^2_{\rm tot}=1.2$, 3.0 and 1.2 and $\chi^2_{\rm tot}=1.3$, 3.0 and 1.6 for the standard continuity prior with and without emission line constraints. Including or excluding emission lines has only little effect on the resulting $\chi^2$ values.

\subsection{Magnification factors}
\label{subsec:magnification}

We use the magnification factors derived by \citet{curti23} for each galaxy (Table~\ref{tab:galaxies}). Briefly, \citet{curti23} adopted the publicly available lens model of \citet{mahler23} to derive the magnification corrections, which combine ancillary \HST with novel \JWST/NIRCam data to better constrain the cluster mass distribution. For each object, \citet{curti23} derived the magnification maps for each target redshift and then obtained the fiducial magnification factors by averaging its value within a 1\arcsec-wide box around the central coordinates of each galaxy. 

Other lens models have been published in the literature \citep[e.g.,][]{pascale22, caminha22}. The two weakly magnified objects (Galaxy ID 06355 and 10612) are robust, but we note that for the highest magnification galaxy ID4590, the different values obtained by the available models can impact the stellar mass and SFR determination for this source by up to $30-50\%$.

\section{Results}
\label{sec:results}

\begin{figure*}
	\includegraphics[width=\textwidth]{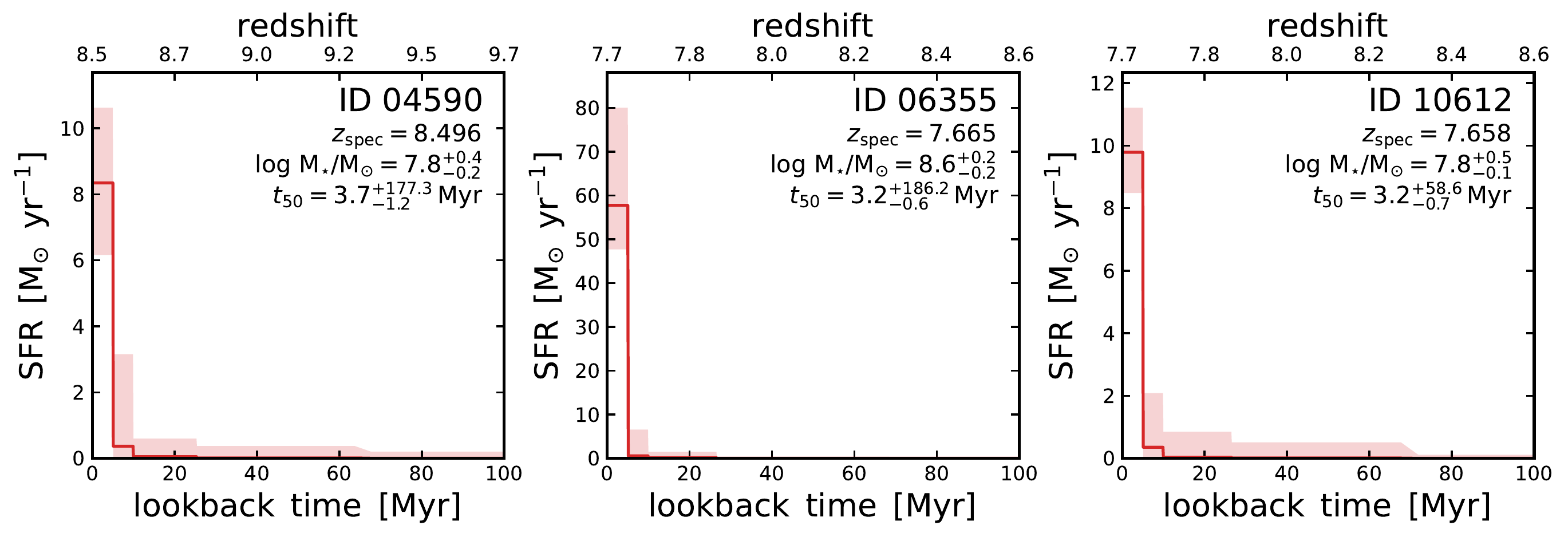}
     \caption{Star-formation histories (SFHs) over the past 100 Myr. The solid red line indicates the median of the posterior distribution inferred with our fiducial model, while the shaded region mark the 16th-84th percentiles. All of the galaxies are consistent with a recent burst of star formation: the SFRs in the past 5 Myr are higher by a factor of $\geq10$ than 20 Myr ago. Importantly, these are not the full SFHs, which extend back to $z=20$.}
    \label{fig:SFH}
\end{figure*}

\begin{figure}
	\includegraphics[width=0.95\linewidth]{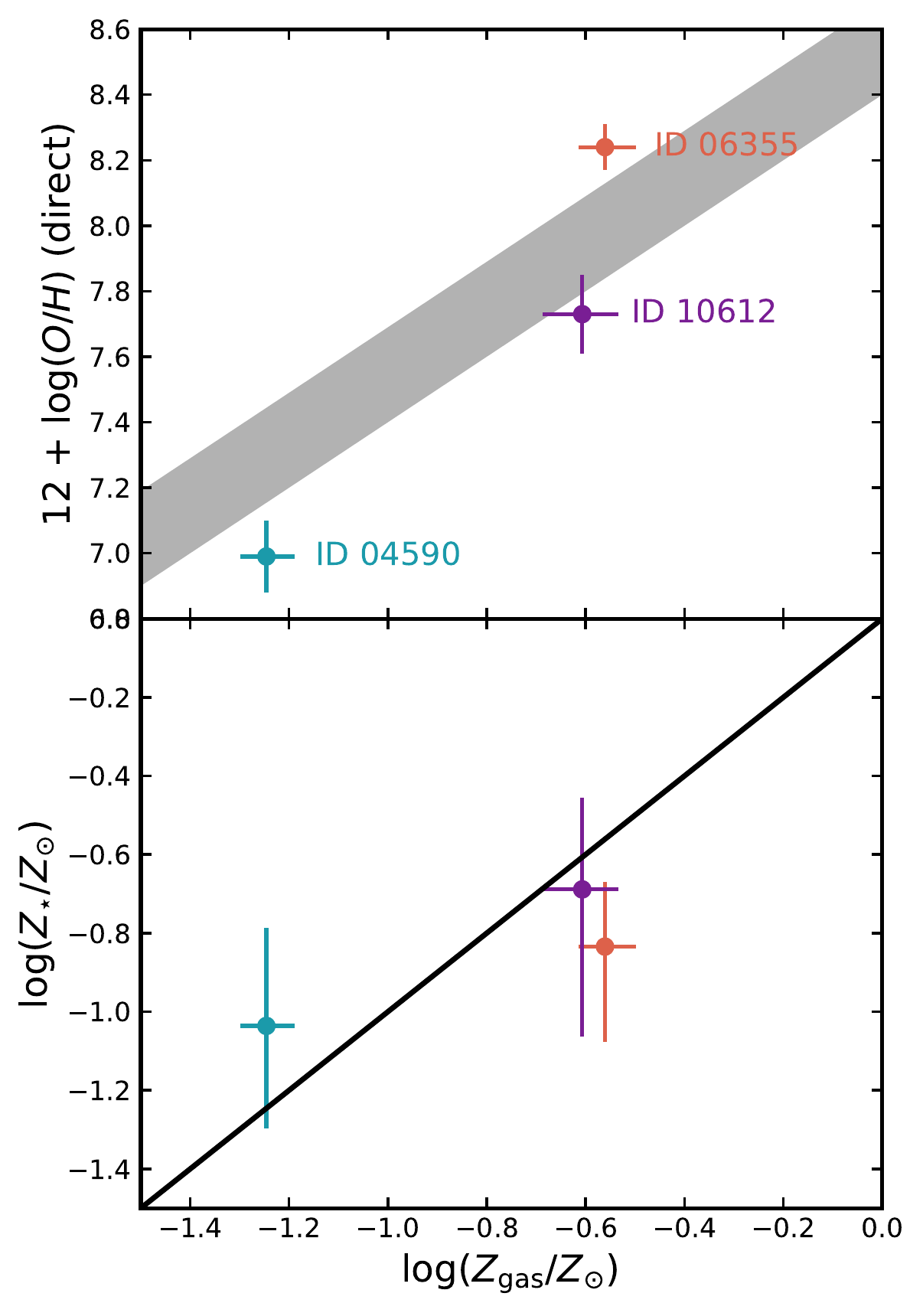}
    \caption{Comparison of the inferred gas-phase metallicity to the direct gas-phase metallicity (top panel) and to the inferred stellar metallicity (bottom panel). The direct gas-phase metallicity is inferred from the [OIII]4364 emission line \citep{curti23}. The grey region in the top panel marks a solar oxygen abundance of $12+\log(\mathrm{O}/\mathrm{H})_{\odot}=8.40-8.69$ \citep{asplund09}. We find a good agreement between the gas-phase metallicity and the directly inferred oxygen abundance. We find a $\mathrm{Z}_{\rm gas}\lesssim\mathrm{Z}_{\star}$ for Galaxy ID 04590, while Galaxy ID 06355 and 10612 have $\mathrm{Z}_{\rm gas}\gtrsim\mathrm{Z}_{\star}$ -- but we note that the stellar metallicity is uncertain.} 
    \label{fig:metallicity_comparison}
\end{figure}

\begin{figure*}
	\includegraphics[width=\textwidth]{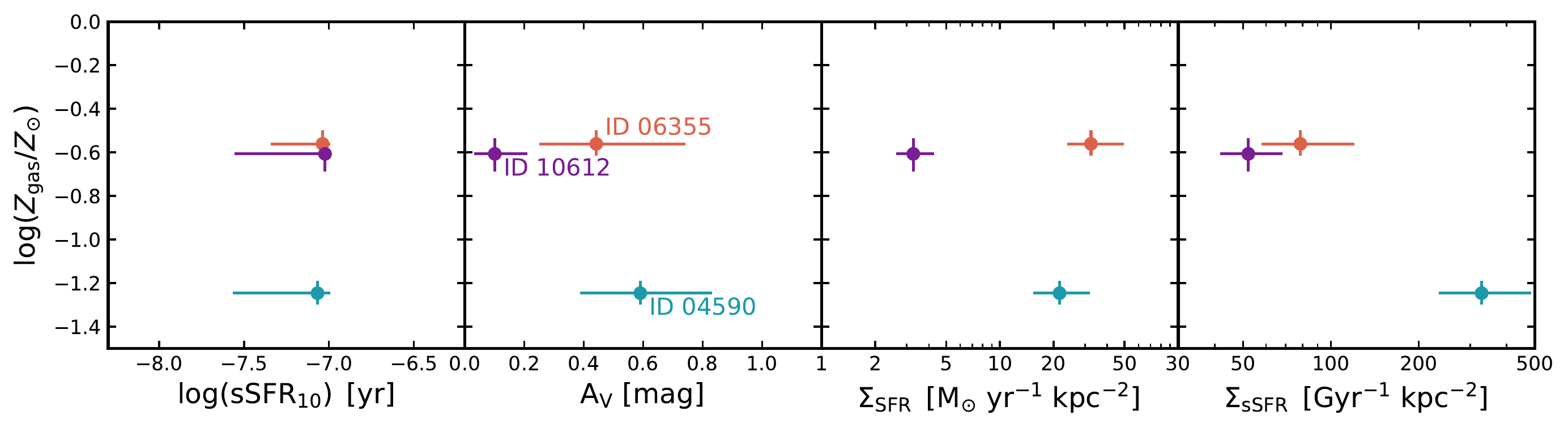}
    \caption{Correlation of the gas-phase metallicity $\mathrm{Z}_{\rm gas}$ with sSFR$_{10}$ ($\mathrm{sSFR}_{10}=\mathrm{SFR}_{10}/M_{\star}$, i.e. averaged over 10 Myr; left panel), dust attenuation A$_{\rm V}$ (middle left panel), SFR surface density ($\Sigma_{\rm SFR}=\mathrm{SFR}_{10}/(2\pi R_{\rm e}^2)$; middle right panel), and sSFR surface density ($\Sigma_{\rm sSFR}=\mathrm{SFR}_{10}/(M_{\star} 2\pi R_{\rm e}^2)$; right panel). There is no clear trend between $\mathrm{Z}_{\rm gas}$ and sSFR$_{10}$ (and also stellar age). Intriguingly, Galaxy ID 04590, which has to lowest $\mathrm{Z}_{\rm gas}$, has an extremely high $\Sigma_{\rm sSFR}$, i.e. is vigorously forming stars in a very compact configuration, indicating that this galaxy is likely experiencing a rapid rising phase of enrichment while being fuelled by accretion of pristine gas.} 
    \label{fig:metallicity_sfh}
\end{figure*}

\begin{figure*}
	\includegraphics[width=\textwidth]{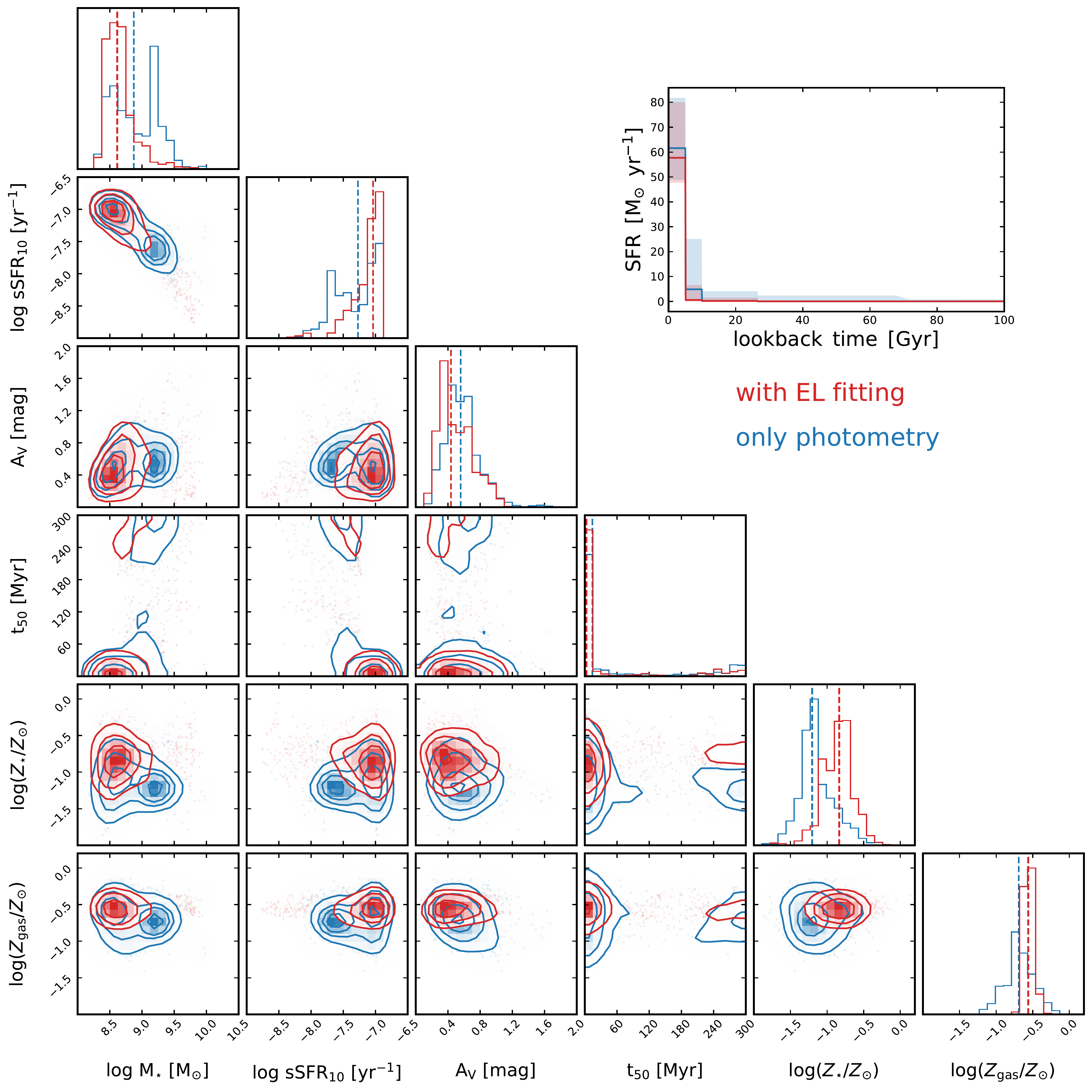}
    \caption{Effect of including emission lines in the fitting for Galaxy ID 06355. The results for including emission lines in the fitting and only fitting the NIRCam photometry are shown in red and blue, respectively. Including emission lines helps with constraining the gas-phase metallicity, stellar age and SFR.} 
    \label{fig:effect_EL}
\end{figure*}

\begin{figure*}
	\includegraphics[width=\textwidth]{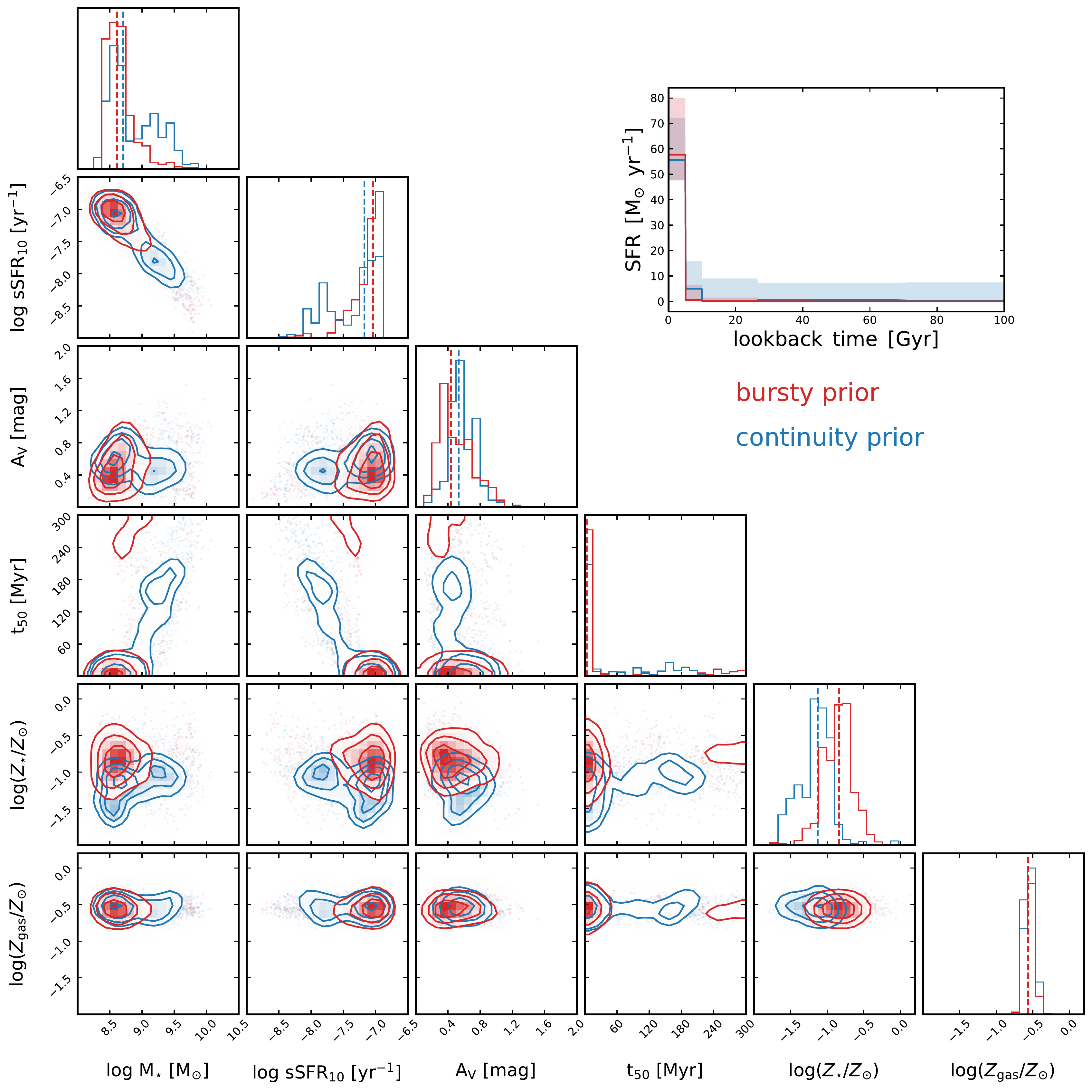}
    \caption{Effect of changing the SFH prior for Galaxy ID 06355. The results adopting the bursty prior and the continuity prior for the SFH are shown in red and blue, respectively (the red histogram is the same as in Figure~\ref{fig:effect_EL}). The continuity prior results in a more constant SFH, which leads to a higher stellar mass, lower sSFR and older age.} 
    \label{fig:effect_sfh_prior}
\end{figure*}

After showing in the previous section that our fits are able to reproduce the observational data, we present and discuss here the key results for our fiducial \texttt{Prospector} run regarding the early stellar mass build-up (Section~\ref{subsec:sfh}) and enrichment (Section~\ref{subsec:enrichment}). We then discuss why self-consistent emission line modelling is important (Section~\ref{subsec:EL}) and what the effects are of changing the SFH prior (Section~\ref{subsec:sfh_prior}). Finally, we close by highlighting several caveats and future improvements (Section~\ref{subsec:caveats}).

\subsection{Stellar mass build-up}
\label{subsec:sfh}

Figure~\ref{fig:SFH} plots the inferred SFHs and also gives the stellar masses and mass-weighted ages. The SFHs are increasing for all three galaxies and have similar stellar masses of about $10^8~M_{\odot}$, with Galaxy ID 04590 and 10612 being the least massive ($M_{\star}\approx10^{7.8}~M_{\odot}$) and and  Galaxy ID 06355 being the most massive galaxy ($M_{\star}\approx10^{8.6}~M_{\odot}$). Galaxy ID 04590 has increased its SFR from $<0.1~\mathrm{M}_{\odot}~\mathrm{yr}^{-1}$ 100 Myr ago to $>5~\mathrm{M}_{\odot}~\mathrm{yr}^{-1}$ in the past 5 Myr, i.e. this galaxy is undergoing a recent burst. This is consistent with its young, mass-weighted age of $t_{50}=4_{-3}^{+177}$ Myr. We note that older ages are still possible and difficult to rule out because of the strong outshining effect due to the ongoing starburst. This young age is consistent with the morphological appearance, which indicates that this source is compact with a size of less than 0.2 kpc. 

Galaxy ID 06355 is going through an even larger burst of star formation. In this case, older ages can largely be ruled out\footnote{This galaxy has the reddest F444W-F356W colour, indicating strong emission lines.} and we inferred the youngest age of this sample with $3_{-1}^{+186}$ Myr. It is interesting to note that this galaxies is more extended than Galaxy ID 4590 and is a clumpy object (Figure~\ref{fig:data6355}), consistent with the idea that mergers or an internal, gravitational instability triggered this intense starburst. Galaxy ID 10612 has an age of $t_{50}=3_{-1}^{+59}$ Myr. The SFH shows a hint of elevated star formation $30-60$ Myr ago. Together with insight that this source is made-up of two components (Figure~\ref{fig:data10612}), it is consistent with a picture where we are witnessing a merger in progress. 

For comparison, \citet{carnall23_ero} fitted the \HST and \JWST photometry with the SED fitting code \texttt{Bagpipes} \citep{carnall18}, inferring stellar masses (corrected for the different assumptions related to magnification) of $\log(M_{\star}/M_{\odot})=7.5_{-0.2}^{+0.2}$, $8.6_{-0.1}^{+0.1}$, and $7.8_{-0.1}^{+0.1}$, and mean stellar ages of $1.3_{-0.3}^{+1.3}$ Myr, $1.3_{-0.3}^{+0.4}$ Myr and $1.2_{-0.2}^{+0.3}$ Myr for Galaxy ID 04590, 06355 and 10612, respectively. Both the stellar masses and ages are consistent within the uncertainties. The ages of \citet{carnall23_ero} are at lower end of our quantiles, but still consistent with our estimates. However, \citet{carnall23_ero} seem to rule out any underlying older component for those galaxies, which contrasts with our estimates. It is not straightforward to pinpoint the cause for this difference, but it most probably has to do with the different SFH prescriptions. \citet{carnall23_ero} assume a simple constant SFH, while we assume a more flexible SFH. In addition, different extraction methods of the photometry could also lead to some differences. Finally, \citet{carnall23_ero} only fit the photometry, while we fit both photometry and emission lines. As shown in Section~\ref{subsec:EL}, we would however expect that this leads to the opposite trend: excluding emission lines leads typically to older ages (Table~\ref{tab:model_variation}).

\subsection{Dust \& enrichment}
\label{subsec:enrichment}

Figure~\ref{fig:metallicity_comparison} investigates our inferred gas-phase and stellar metallicities. In particular, the top panel compares our gas-phase metallicities ($\mathrm{Z}_{\rm gas}$) from the \texttt{Prospector} SED fitting with the ones from \citet{curti23}, which used the direct $T_{\rm e}$ method enabled by the detection of [OIII]4363. Encouragingly, we find very good agreement between the two approaches, confirming the rather low gas-phase metallicity of Galaxy ID 04590 with $\mathrm{Z}_{\rm gas}=0.06~\mathrm{Z}_{\odot}$. 

We find that the gas-phase and stellar metallicities are marginally consistent, though we emphasise that the inferred stellar metallicities are uncertain and that the elements that the gas-phase and stellar metallicity trace are different. Specifically, gas-phase metallicity traces the Oxygen abundance, while stellar metallicity traces mostly Iron. Therefore, one would expect an enhancement in the gas-phase metallicity of $\sim2-5$ times compared with the stellar metallicity \citep{strom22, arellano-cordova22}. We find that Galaxy ID 04590 has $\mathrm{Z}_{\rm gas}\lesssim\mathrm{Z}_{\star}$, while the other two have $\mathrm{Z}_{\rm gas}\gtrsim\mathrm{Z}_{\star}$.

Figure~\ref{fig:metallicity_sfh} correlates the metallicities with the inferred stellar population and morphological parameters, including sSFR$_{10}$ (averaged over past 10 Myr), dust attenuation $\mathrm{A}_{\rm V}$, SFR surface density $\Sigma_{\rm SFR}$, and sSFR surface density $\Sigma_{\rm sSFR}$. We do not find any convincing trend for $\mathrm{Z}_{\rm gas}$ with either sSFR$_{10}$ (as well as $t_{50}$). Interestingly, the two galaxies (Galaxy ID 04590 and ID 06355) with an elevated $A_{\rm V}$ and $\Sigma_{\rm SFR}$ have very different gas-phase metallicities. The possible cause from this can be found in the right panel of Figure~\ref{fig:metallicity_sfh}: these two galaxies  have very different sSFR surface densities with $\Sigma_{\rm sSFR}=328_{-94}^{+156}~\mathrm{Gyr}^{-1}~\mathrm{kpc}^{-2}$ (Galaxy ID 04590) and $79_{-21}^{+42}~\mathrm{Gyr}^{-1}~\mathrm{kpc}^{-2}$ (Galaxy ID 06355), given by the fact that Galaxy ID 04590 is a factor of $2-3$ times more compact and lower in stellar mass than Galaxy ID 06355 (Figures~\ref{fig:data4590} and \ref{fig:data6355}). Galaxy ID 04590 is vigorously forming stars in a very compact configuration. This is consistent with a scenario in which Galaxy ID 04590 is undergoing rapid accretion of pristine gas, which leads to a steeply rising SFH and a high (s)SFR surface density. Despite having a low gas-phase metallicity, the high $A_{\rm V}$ can be explained by its compactness (i.e. high gas surface density\footnote{The dust-star geometry plays an important role in setting the attenuation \citep[e.g.,][]{zuckerman21}.}). Consistent with this rapid inflow of pristine gas is the tentative evidence of $\mathrm{Z}_{\rm gas}\lesssim\mathrm{Z}_{\star}$. Furthermore, the source just left of Galaxy ID 04590 in Fig.~\ref{fig:data4590} might be part of this accretion stream. The two other galaxies (Galaxy ID 06355 and 10612) with higher gas-phase metallicity show more complex multi-component morphologies on kpc scale, indicating that their recent increase in SFR is driven by mergers or internal, gravitational instabilities. This is consistent with the FMR analysis of \citet{curti23}, where both Galaxy ID 06355 and 10612 are roughly consistent with the FMR at lower redshifts, while Galaxy ID 04590 deviates significantly, implying it being far from the smooth equilibrium between gas flows, star formation and metal enrichment in place at later epochs.

\subsection{The constraining power of emission lines}
\label{subsec:EL}

From a pure Bayesian modelling approach, we want to include information to constrain the posterior distribution as tightly as possible. We therefore want to fit both the broad-band photometry and the emission lines. The emission lines themselves include crucial information about the gas properties (e.g., metallicity, temperature, and density) and  ionizing sources (stars and other sources), including the most recent SFR.  

Figure~\ref{fig:effect_EL} shows the posterior distribution of the fits including both photometry and emission lines (red; fiducial) and only photometry (blue) for Galaxy ID 06355. The results are summarised for all objects in Table~\ref{tab:model_variation}. An obvious, straightforward difference is that the gas-phase metallicity is basically unconstrained when fitting only the photometry. In addition, since the relative line ratios are sensitive to the dust attenuation (in particular Balmer line series), the dust attenuation is more tightly constrained when including emission lines, which also affects the stellar age constraint via the rest-UV (although some degeneracy remains due to the flexible dust attenuation law). Directly related, another large difference can be seen for the inferred SFHs: without emission line constraints, the inferred stellar ages are older since the most recent SFH is less well constrained. This then also leads to larger stellar masses (up to $\sim0.4$ dex in the case of Galaxy ID 04590), indicating that most inferred properties of the galaxies are affected by whether emission lines are included or not in the fits.

\subsection{Effects of the SFH prior}
\label{subsec:sfh_prior}

Since all these galaxies seem to be undergoing a recent burst in star formation (Figure~\ref{fig:SFH}), young stellar populations dominate the SED and outshine older stellar populations. Therefore, a significant amount of stellar mass could be hidden and is difficult to rule out, which implies that the prior on the SFH is expected to play an important role. As mentioned in Section~\ref{subsec:runs}, we run \texttt{Prospector} with two SFH priors: our fiducial ``bursty'' prior, which allows for large variations between adjacent time bins, and the standard ``continuity'' prior, which down-weights extreme bursts.

Figure~\ref{fig:effect_sfh_prior} shows the effect of the SFH prior on key inferred quantities, including the SFH, stellar mass, sSFR, dust attenuation, age, stellar metallicity and gas-phase metallicity. The results from the bursty and standard continuity prior are shown in red and blue, respectively. The same observational data, i.e. both photometry and spectroscopy, are fitted (``EL \& bursty'' and ``EL \& continuous'' runs in Table~\ref{tab:model_variation}). 

As expected, the SFH prior affects the inferred SFH: the continuous SFH prior leads to more stellar mass being formed early and a less steep rise in recent time. This leads to overall older ages, larger stellar masses and lower sSFRs than what is inferred with the bursty prior. Specifically, as tabulated in Table~\ref{tab:model_variation}, the stellar ages increase from $\lesssim10$~Myr to $\gtrsim100$ Myr for Galaxy ID 4590 and 10612, highlighting that it is difficult to rule out early star formation. Importantly, the $t_{50}$ posterior distributions of the two runs are overlapping, i.e. they are consistent with each other. There is also a big impact on the inferred stellar masses: they increase by up to 0.6 dex for those two galaxies. 

The stellar age constraints come from the rest-UV and 4000\AA-break \citep[e.g.,][]{conroy13_rev}. Since we are using a flexible dust attenuation law (Section~\ref{subsec:prospector}), the rest-UV has only little constraining power. The 4000\AA-break for our three galaxies at $z\sim7-9$ is probed by the broad-band photometry, but -- as shown in Figure~\ref{fig:sed} -- there is a degeneracy with the strength of the emission lines. Specifically, both a strong continuum break and strong emission lines can lead to a boost of the longer-wavelength filter relative to the shorter-wavelength one. But why are we not able to do better when including emission lines in the fitting? The problem is that our nuisance parameter, which rescales the emission line fluxes, is degenerate with the strength of the 4000\AA-break, i.e. our emission line measurements mainly constrain the gas-phase metallicity and the dust attenuation, which in second order leads to a tighter constraint on the stellar ages via the rest-UV. 

Further improvements (and solving the degeneracy with slit losses) can be achieved by exactly assessing the slit position (and performing spatially resolved SED modelling) and by directly probing emission lines with the photometry, which can be achieved with narrow and medium bands \citep[e.g.,][]{roberts-borsani21_jwst, tacchella22_highz}. The medium bands on/off the lines would allow the measurement of the 4000\AA-break more directly.

\subsection{Caveats and outlook}
\label{subsec:caveats}

We list here the most important caveats in our analysis, which we hope to improve upon in the future. From the Bayesian SED modelling perspective, our goal is to include as much information as possible that can be described by our physical model. This means we want to fit both the NIRCam photometry and the NIRSpec emission line constraints. One fundamental challenge with this is that the photometry and spectroscopy could probe different regions of the galaxy. Specifically, the photometry attempts to include the total emission of the galaxy, while the spectroscopy only captures the light that is within the shutter of the NIRSpec MSA. Slit losses affect most spectroscopic observations, but because not every object can be aligned perfectly with the NIRSpec MSA, slit losses may be a larger issue for these observations. Besides this technical challenge\footnote{Another technical challenge is that the level-3 data products of the current STScI pipeline lead to unphysical ratios between the Balmer lines (e.g. H$\upgamma$/H$\upbeta$; see, e.g., \citealt{schaerer22, curti23, trump23}) -- one key reason why we used the reprocessed data from \citet{curti23}. This challenge will hopefully be resolved in the near future.}, other observational and physical effects can also lead to a mismatch between the photometry and spectroscopy. On the observational side, differential magnification across the source can lead to boosting of flux of certain galactic sub-regions (typically of more compact regions), leading to a bias between the photometry and spectroscopy (as well as of certain emission lines). On the physical side, we assume in this work that the emission lines (and the nebular continuum) are powered by ionizing photons of the stellar populations, without taking into account LyC absorption by dust of LyC escape, which could be important, in particular for objects in the starburst phase \citep[e.g.,][]{tacchella22_Halpha}. 

In this work, we addressed these technical, observational and astrophysical issues by fitting for a nuisance parameter, $f_{\rm scale}$, which scales all the emission lines by a constant factor, mainly intended to account for slit loss. We find $f_{\rm scale}$ to be $0.57_{-0.07}^{+0.07}$, $0.50_{-0.07}^{+0.11}$, and $0.31_{-0.04}^{+0.05}$ for Galaxy ID 04590, 06355 and 10612, respectively. If this rescaling is ignored, we find extremely high values for the stellar metallicities ($\mathrm{Z}_{\star}>\mathrm{Z}_{\odot}$). This highlights that this factor is important when performing the fitting. Our inferred $f_{\rm scale}$ are larger than calculated from slit loss, implying that 31\%, 21\% and 30\% need be explained via other means, for example Lyman continuum absorption by dust or escape. Furthermore, working with unlensed galaxies will resolve the issue with the differential magnification bias. 

Related to this is the assumption in our approach that stellar populations are driving the emission lines. Firstly, active galactic nuclei (AGN) and other non-stellar sources could power the emission lines as well as contribute to the rest-UV emission. Within \texttt{Prospector}, only the rest-frame mid-infrared emission of AGN can be modelled (i.e. dusty torus emission). There are on-going efforts to expand upon this. Secondly, a recent analysis by \citet{katz23} shows that the inclusion of high-mass X-ray binaries or a high cosmic ray background in addition to a young, low-metallicity stellar population can provide the additional heating necessary to explain the observed high [OIII]4364/[OIII]5007 ratio while remaining consistent with other observed line ratios. However, because it is challenging photoionization models to predict [OIII]4364 \citep[e.g.,][]{dors11}, we mask this line. More generally, it is notoriously difficult to distinguish between AGN and star formation as an ionization source at low metallicity. Interestingly, Galaxy ID 6355 appears to have an AGN contribution (narrow-line AGN given [NeIV]2422,2424 is detected; \citealt{brinchmann22}). We believe that our results are robust even for this object given that such high ionization lines can be obtained by adding an Xray source without strongly affecting the other lines \citep{katz23}.

For completeness, we also mention the caveats related to stars. In particular, in this work we assume a solar abundance pattern scaled by the metallicity. However, there is an indication that at higher redshifts (for $z\approx2$ results see \citealt{strom22}) galaxies are more $\alpha$-enhanced (but see \citealt{arellano-cordova22} for a recent analysis of this for those three galaxies, finding no evolution in this $\alpha$-element ratio). This is also theoretically expected given that some of the high-$z$ galaxies are old enough to have seen enrichment from intermediate-mass stars, but are still young enough that Type Ia supernovae have not had time to contribute significantly to their enrichment \citep[e.g.,][]{kriek16}. We further assume the MIST stellar models \citep{choi17}, which include rotation, but not binarity. Investigating the effects the binary-based stellar model \citep[e.g.,][]{eldridge17} or varying the IMF is beyond the scope of this work.

Finally, this work investigates three galaxies, the only three $z>7$ galaxies that have both NIRCam and NIRSpec observations. Obviously, we cannot draw strong conclusions regarding the whole galaxy population from three objects. Although we think that galaxies have mostly increasing SFHs at these early times \citep[e.g.,][]{tacchella18, endsley21_ew}, the substantial bursts found in these galaxies might be a selection effect and not representative of all galaxies at $z\sim7-9$ \citep[e.g.,][]{looser23}. Therefore, we stress that the main purpose of this work is to deliver interesting insights into the properties of those three galaxies by combining NIRSpec and NIRCam data, discuss technical but important details for doing this, and discuss how priors play an important role in driving some of the results.

\section{Conclusions}
\label{sec:conclusions}

We present a careful ISM and stellar population analysis of the ERO NIRCam and NIRSpec data of three $z=7.6-8.5$ galaxies in the SMACS cluster field. These three galaxies have diverse morphologies (Figures~\ref{fig:data4590}-\ref{fig:data10612}), from being compact (Galaxy ID 04590) to being made up of several components (at least two components in Galaxy ID 10612 and four components in Galaxy ID 06355). We perform the photometry with the new photometry tool \texttt{forcepho}, which is a Bayesian model-fitting program that simultaneously fits multiple PSF-convolved S\'{e}rsic profiles to all filters. 

Within the \texttt{Prospector} framework, we fit a 13-parameter model to NIRCam photometry and the NIRSpec emission lines. This physical SED model includes a flexible SFH, a multi-component dust model including a variable dust attenuation law, different gas-phase and stellar metallicities, a free ionization parameter for the nebular emission and a nuisance parameter that scales the emission line fluxes. The latter parameter is important to account for possible slit-loss effects and physical effects such as LyC dust absorption and escape. We find that this factor is important: model-based emission line fluxes need to be rescaled by factors of $0.3-0.6$, caused both by slit loss and astrophysical reasons (for example escape of ionising radiation). Overall, we find that we are able to reproduce the photometry and emission line measurements (Figure~\ref{fig:sed}). An exception is the emission line [OIII]4364, which is brighter than predicted by  our modelling, which could be attributed to non-stellar sources powering this line, to a conservative \texttt{cloudy} grid (too low ionization parameter values).

We infer for all three galaxies rising SFHs and stellar masses of $M_{\star}\approx10^{8}~M_{\odot}$ (Figure~\ref{fig:SFH} and Table~\ref{tab:galaxies}). These galaxies are all young with mass-weighted ages of $t_{50}=3-4$ Myr. However, we find indications for underlying older stellar populations, implying the SFHs extend at least over several tens of Myr. We emphasise that the SFHs, stellar masses and stellar ages depend on the adopted SFH prior (Figure~\ref{fig:effect_sfh_prior} and Table~\ref{tab:model_variation}): assuming a bursty SFH prior leads to younger, lower-mass galaxies, which is consistent with previous studies \citep{tacchella22_highz, whitler23_sfh}. Emission lines are helpful to constrain mainly the gas-phase metallicity and ionization parameter, and only have a second order effect on the inferred SFHs and ages (Figure~\ref{fig:effect_EL} and Table~\ref{tab:model_variation}). This is because of the aforementioned rescaling of the emission lines, i.e. we are fitting for relative emission line strength and not their absolute strength. However, we still find that emission lines help with constraining the SFHs because they pin down the dust attenuation parameters, which then in turn constrains the rest-UV emission -- another important age indicator. 

Focusing on metallicity, our SED fitting approach (with masking the [OIII]4364 line) delivers gas-phase metallicities consistent with the direct method from the [OIII]4364 line from \citet[][Figure~\ref{fig:metallicity_comparison}]{curti23}. We find no convincing trend between the gas-phase metallicity and the total sSFR (and stellar age) of the galaxies (Figure~\ref{fig:metallicity_sfh}). However, Galaxy ID 04590 with the lowest gas-phase metallicity shows the highest star-formation intensity as measured by the sSFR surface density $\Sigma_{\rm sSFR}$ ($\Sigma_{\rm sSFR}\approx330~\mathrm{Gyr}^{-1}~\mathrm{kpc}^{-2}$), i.e. this galaxy has very high SFR surface ($\Sigma_{\rm SFR}\approx22~\mathrm{M}_{\odot}~\mathrm{yr}^{-1}~\mathrm{kpc}^{-2}$) density for its stellar mass ($M_{\star}\approx10^8~\mathrm{M}_{\odot}$), indicating that this galaxy is vigorously forming stars in a very compact configuration. This is consistent with a scenario in which Galaxy ID 04590 is undergoing rapid accretion of pristine gas, which leads to a steeply rising SFH and a high (s)SFR surface density. Despite having a low gas-phase metallicity, the high $A_{\rm V}$ can be explained by its compactness (i.e. high gas surface density). Consistent with this rapid inflow of pristine gas is the tentative evidence of $\mathrm{Z}_{\rm gas}\lesssim\mathrm{Z}_{\star}$. The two other galaxies (Galaxy ID 06355 and 10612) with higher gas-phase metallicity show more complex multi-component morphologies on kpc scales, indicating that their recent increase in SFR is driven by mergers or internal, gravitational instabilities. 

In summary, our work highlights the great potential for combining photometric with spectroscopic \JWST data to study early galaxy formation, such as the JADES survey \citep{rieke19, bunker23, cameron23, curtis-lake23, robertson23, saxena23, tacchella23}.

\section*{Acknowledgements}

We thank the referee for their helpful comments that improved the manuscript. We are grateful to Pierre Ferruit, Peter Jakobsen, and Nora L\"{u}tzgendorf for sharing their expertise on NIRSpec and the processing of its unique data. We thank Matt Auger for discussing magnification effects.

This work is based on observations made with the NASA/ESA/CSA James Webb Space Telescope. The data were obtained from the Mikulski Archive for Space Telescopes at the Space Telescope Science Institute, which is operated by the Association of Universities for Research in Astronomy, Inc., under NASA contract NAS 5-03127 for \JWST. These observations are associated with programme \#2736. The authors acknowledge the SMACS ERO team led by Klaus Pontoppidan for developing their observing program with a zero-exclusive-access period.

This work is supported by JWST/NIRCam contract to the University of Arizona, NAS5-02015, by the Science and Technology Facilities Council (STFC), by the European Research Council (ERC) Advanced Grant 695671 ``QUENCH'', by the ERC Advanced Grant INTERSTELLAR H2020/740120, and by the ERC Advanced Grant 789056 ``FirstGalaxies''.

\section*{Data Availability}

Derived data (including the reduced NIRCam images, photometry and \texttt{Prospector} posterior distributions) supporting the findings of this study are available from the corresponding author ST on request. Fully reduced NIRSpec spectra are publicly available at \url{https://doi.org/10.5281/zenodo.6940561}.


\bibliographystyle{mnras}



\appendix

\section*{Affiliations}
\noindent
{\it
$^{1}$Kavli Institute for Cosmology, University of Cambridge, Madingley Road, Cambridge, CB3 0HA, UK\\
$^{2}$Cavendish Laboratory, University of Cambridge, 19 JJ Thomson Avenue, Cambridge, CB3 0HE, UK\\
$^{3}$Center for Astrophysics $\vert$ Harvard \& Smithsonian, 60 Garden Street, Cambridge, MA 02138, USA\\
$^{4}$Department of Astronomy and Astrophysics, University of California, Santa Cruz, 1156 High Street, Santa Cruz, CA 95064, USA\\
$^{5}$Scuola Normale Superiore, Piazza dei Cavalieri 7, I-56126 Pisa, Italy\\
$^{6}$AURA for the European Space Agency, Space Telescope Science Institute, 3700 San Martin Drive, Baltimore, MD 21218, USA\\
$^{7}$Department of Physics and Astronomy, University College London, Gower Street, London, WC1E 6BT, UK\\
$^{8}$Department for Astrophysical and Planetary Science, University of Colorado, Boulder, CO 80309, USA\\
$^{9}$Department of Astronomy and Astrophysics, University of California, Santa Cruz, 1156 High Street, Santa Cruz, CA 95064 USA\\
$^{10}$Kavli Institute for Particle Astrophysics and Cosmology and Department of Physics, Stanford University, Stanford, CA 94305, USA\\
$^{11}$NSF's National Optical-Infrared Astronomy Research Laboratory, 950 North Cherry Avenue, Tucson, AZ 85719, USA\\
$^{12}$Steward Observatory, University of Arizona, 933 N. Cherry Avenue, Tucson, AZ 85721, USA\\
$^{13}$Centro de Astrobiolog\'{i}a, (CAB, CSIC–INTA), Departamento de Astrof\'{i}sica, Cra. de Ajalvir Km. 4, 28850 – Torrej\'{o}n de Ardoz, Madrid, Spain\\
$^{14}$European Space Agency, ESA/ESTEC, Keplerlaan 1, 2201 AZ Noordwijk, NL\\
$^{15}$Cosmic Dawn Center, Niels Bohr Institute, University of Copenhagen, Radmandsgade 62, 2200 Copenhagen N, Denmark\\
$^{16}$Jodrell Bank Centre for Astrophysics, Department of Physics and Astronomy, School of Natural Sciences, The University of Manchester, Manchester, M13 9PL, UK\\
$^{17}$Department of Physics, University of Oxford, Denys Wilkinson Building, Keble Road, Oxford OX1 3RH, UK\\
$^{18}$University of Massachusetts Amherst, 710 North Pleasant Street, Amherst, MA 01003-9305, USA\\
$^{19}$Observatoire de Gen\`{e}ve, Universit\'{e} de Gen\`{e}ve, Chemin Pegasi 51, 1290 Versoix, Switzerland\\
$^{20}$NRC Herzberg, 5071 West Saanich Rd, Victoria, BC V9E 2E7, Canada
}




\bsp	
\label{lastpage}
\end{document}